# Thermodynamics of ketone + amine mixtures. Part I. Volumetric and speed of sound data at (293.15, 298.15 and 303.15) K for 2-propanone + aniline, + *N*-methylaniline, or + pyridine systems

IVÁN ALONSO, VÍCTOR ALONSO, ISMAEL MOZO, ISAÍAS GARCÍA DE LA FUENTE, JUAN ANTONIO GONZÁLEZ[*], AND JOSÉ CARLOS COBOS

G.E.T.E.F., Departamento de Física Aplicada, Facultad de Ciencias, Universidad de Valladolid, 47071 Valladolid, Spain,

*e-mail: jagl@termo.uva.es; Fax: +34-983-423136; Tel: +34-983-423757

**Abstract**

Densities, $\rho$, and speeds of sound, $u$, of 2-propanone + aniline, + *N*-methylaniline, or + pyridine systems have been measured at (293.15, 298.15 and 303.15) K and atmospheric pressure using a vibrating tube densimeter and sound analyser Anton Paar model DSA-5000. The $\rho$ and $u$ values were used to calculate excess molar volumes, $V^E$, and the excess functions at 298.15 K for the speed of sound, $u^E$, the thermal expansion coefficient, $\alpha_P^E$, and for the isentropic compressibility, $\kappa_S^E$ at 298.15 K. $V^E$ and $\kappa_S^E$ are both negative magnitudes and increase in the same sequence: aniline < *N*-methylaniline < pyridine. The $V^E$ and $\kappa_S^E$ curves are shifted towards higher mole fractions of 2-propanone. The data haven interpreted assuming strong acetone-amine interactions, and weak structural effects.

## Introduction

Amides, amino acids, peptides and their derivatives are of interest because they are simple models in biochemistry. *N*-methylformamide possesses the basic ($-CO$) and acidic ($-NH$) groups of the very common, in nature, peptide bond.[1] As a matter of fact, proteins are polymers of amino acids linked to each other by peptide bonds. Cyclic amides are also of importance due to they are related to structural problems in biochemistry. Consequently, the understanding of liquid mixtures involving the amide functional group is necessary as a first step to a better knowledge of complex molecules of biological interest.[2] So, the aqueous solution of dimethylformamide is a model solvent representing the environment of the interior of proteins. Amides have many practical applications. For example, dimethylformamide and *N*-methylpyrrolidone are used as highly selective extractants for the recovery of aromatic and saturated hydrocarbons from petroleum feedstocks,[3] and $\varepsilon$-caprolactam is used for the production of nylon 6, which is a polycaprolactam formed by ring-opening polymerization. The study of alkanone + amine mixtures, which contain the carbonyl and amine groups in separate molecules, is then pertinent in order to gain insight of amide solutions. In this first article, we report densities, speeds of sound and excess molar volumes at (293.15 K, 298.15 K and 303.15) K, and $\kappa_S^E$, $u^E$ and $\alpha_P^E$ at 298.15 K for 2-propanone + aniline, + *N*-methylaniline, or + pyridine mixtures.

## Experimental

*Materials.* 2-Propanone (67-64-1) and *N*-methylaniline (100-61-8) were from Fluka and aniline (62-53-3) and pyridine (110-86-1) were from Riedel de Haën and used without further purification. Their purity, expressed in mass fraction, was: ≥ 0.995, ≥ 0.98, ≥ 0.995 and ≥ 0.995, respectively. The $\rho$ and $u$ values of the pure liquids are in good agreement with those from the literature (Table 1).

*Apparatus and procedure.* Binary mixtures were prepared by mass in small vessels of about 10 cm$^3$. Caution was taken to prevent evaporation, and the error in the final mole fraction is estimated to be less than ± 0.0001. Conversion to molar quantities was based on the relative atomic mass table of 2006 issued by IUPAC.[4]

The densities and speeds of sound of both pure liquids and of the mixtures were measured using a vibrating-tube densimeter and a sound analyser, Anton Paar model DSA-5000, automatically thermostated within ± 0.01 K. The calibration of the apparatus was carried out with deionised double-distilled water, hexane, heptane, octane, isooctane, cyclohexane and benzene, using $\rho$ values from the literature.[5-7] The accuracy for the $\rho$ and $u$ measurements are ± 1•10$^{-2}$ kg•m$^{-3}$ and ± 0.1 m•s$^{-1}$, respectively, and the corresponding precisions are ± 1•10$^{-3}$ kg•m$^{-3}$ and ± 0.01 m•s$^{-1}$. The experimental technique was checked by determining $V^E$ and $u$ of the standard mixtures: (cyclohexane + benzene) at the temperatures (293.15, 298.15 and 303.15) K and cyclohexane + hexane and 2-ethoxyethanol + heptane at 298.15 K. Our results agree well with published values.[8-11] The accuracy in $V^E$ is believed to be less than $\pm(0.01|V_{max}^E|+0.005)$ cm³•mol⁻¹, where $|V_{max}^E|$ denotes the maximum experimental value of the excess molar volume with respect to the mole fraction. The accuracy of the deviations of $u$ from the ideal behaviour is estimated to be 0.3 m•s⁻¹.

**Equations**

The thermodynamic properties for which values are derived most directly from the experimental measurements are the density, $\rho$, the molar volume, $V$, the coefficient of thermal expansion, $\alpha_P = -\frac{1}{\rho}\left(\frac{\partial \rho}{\partial T}\right)_P$ and the isentropic compressibility, $\kappa_S$. In this work, $\alpha_P$ values were obtained from a linear dependence of $\rho$ with $T$. Assuming that the absorption of the acoustic wave is negligible, $\kappa_s$ can be calculated using the Newton-Laplace's equation:

$$\kappa_S = \frac{1}{\rho u^2} \tag{1}$$

For an ideal mixture at the same temperature and pressure than the system under study, the values $F^{id}$ of the thermodynamic property, $F$, are calculated using the equations:[8,12]

$$F^{id} = x_1 F_1 + x_2 F_2 \qquad (F = V, C_P) \tag{2}$$

and

$$F^{id} = \phi_1 F_1 + \phi_2 F_2 \qquad (F = \alpha_P; \kappa_T) \tag{3}$$

where $C_p$ is the isobaric heat capacity, $\phi_i = \frac{x_i V_i}{V^{id}}$ the volume fraction, $\kappa_T$, the isothermal compressibility, and $F_i$, the $F$ value of component i, respectively. For $\kappa_S$ and $u$, the ideal values are calculated according to:[12]

$$\kappa_S^{id} = \kappa_T^{id} - \frac{TV^{id}\alpha_P^{id2}}{C_P^{id}} \tag{4}$$

and

$$u^{id} = \left(\frac{1}{\rho^{id} \kappa_S^{id}}\right)^{1/2} \qquad (5)$$

where $\rho^{id} = (x_1 M_1 + x_2 M_2)/V^{id}$ ($M_i$, molecular mass of the i component). In this work, we have determined the excess functions:

$$F^E = F - F^{id} \qquad (6)$$

**Results and Discussion**

Table 2 lists values of densities, calculated $V^E$ and of $u$ vs. $x_1$, the mole fraction of the 2-propanone. Table 3 contains the derived quantities $\kappa_S^E$, $u^E$ and $\alpha_P^E$. The data were fitted by unweighted least-squares polynomial regression to the equation:

$$F^E = x_1(1-x_1)\sum_{i=0}^{k-1} A_i(2x_1 - 1)^i \qquad (7)$$

where $F$ stands for the properties cited above. The number of coefficients $k$ used in eq. (7) for each mixture was determined by applying an F-test[13] at the 99.5 % confidence level. Table 4 lists the parameters $A_i$ obtained in the regression, together with the standard deviations $\sigma$, defined by:

$$\sigma\left(F^E\right) = \left[\frac{1}{N-k}\sum\left(F_{cal}^E - F_{exp}^E\right)^2\right]^{1/2} \qquad (8)$$

where $N$ is the number of direct experimental values. Results on $V^E$ and $\kappa_S^E$ are shown graphically in Figures 1 and 2. No data have been encountered in the literature for comparison.

Hereafter, we are referring to values of the excess molar properties at equimolar composition and 298.15 K.

Mixtures of 2-propanone, aniline, N-methylaniline or pyridine with a given alkane are characterized by strong interactions between the corresponding polar molecules. This is supported by the following features. i) The mentioned systems show miscibility gaps in such way that the coexistence curves of the liquid-liquid equilibria have an upper critical solution temperature (UCST). For the aniline + hexane mixture, the UCST is 342.7 K[14] and is 343.11 K[15] for the heptane solution. In the case of pyridine systems, the UCSTs are 252.2 K and 255.2 K, for the mixtures with hexane and heptane, respectively.[16] If one takes into account, that aniline, N-methylaniline and pyridine are primary, secondary and tertiary amines, respectively, and that aniline and N-methylaniline are self-associated via H-bonds, the critical temperatures given above reveal that the strength of amine-amine interactions decrease in the sequence: aniline > *N*-methylaniline > pyridine. For the 2-propanone + heptane mixture, UCST = 245.22 K.[17] (ii) Large excess molar enthalpies, $H^E$, as it is indicated by the following values for heptane systems: $H^E$ (pyridine)[18] = 1735 J•mol$^{-1}$ and $H^E$ (2-propanone)[19] = 1704 J•mol$^{-1}$. (iii) Positive excess molar volumes, $V^E$, which reveals that the disruption upon mixing of amine-amine or ketone-ketone interactions are predominant over other effects which contribute negatively to $V^E$ (structural effects arising from interstitial accommodation of one component into the other and free volume effects). For example, $V^E$ (pyridine + heptane)[20] = 0.2657 cm$^3$•mol$^{-1}$ and $V^E$ (2-propanone + heptane)[21] = 1.130 cm$^3$•mol$^{-1}$.

In addition to structural effects, negative $V^E$ values of binary mixtures may be attributed to strong chemical or specific interactions between unlike molecules. For the investigated systems, the negative $V^E$ values determined here may be due to a large extent to the mentioned interactions. The value $V^E$ (2-propanone + aniline) = −1.183 cm$^3$•mol$^{-1}$ is consistent with the large negative $H^E$ of this solution[22] (−1224 J•mol$^{-1}$), which reveals strong acetone-aniline interactions. The strength of such interactions may be roughly estimated as follows. Let us denote the positive contributions to $H^E$ from the breaking of the 2-propanone-2-propanone and aniline-aniline interactions by $\Delta H_{\text{CO-CO}}$ and $\Delta H_{\text{N-N}}$, respectively, and by $\Delta H_{\text{N-CO}}$, the negative contributions from the creation of the aniline-2-propanone interactions. Taking into account these contributions to $H^E$, we can write:

$$H^E = \Delta H_{\text{CO-CO}} + \Delta H_{\text{N-N}} + \Delta H_{\text{N-CO}} \qquad (9)$$

This equation has been widely applied.[23,24] It can be extended to $x_1 \rightarrow 0$ [25] to evaluate $\Delta H_{\text{(N-CO)bond}}$, the strength of the H-bonds between aniline and 2-propanone. In such case, $\Delta H_{\text{CO-CO}}$ and $\Delta H_{\text{N-N}}$ can be replaced by $H_1^{E,\infty}$ (partial excess molar enthalpy at infinite dilution of the first component) of 2-propanone or aniline + heptane systems. So,

$$\Delta H_{\text{(N-CO)bond}} = H_1^{E,\infty}(\text{aniline} + 2-\text{propanone})$$
$$-H_1^{E,\infty}(\text{aniline} + \text{heptane}) - H_1^{E,\infty}(2-\text{propanone} + \text{heptane}) \qquad (10)$$

The data used in this work are: $H_1^{E,\infty}$ (2-propanone heptane)[26] = 9.84 kJ•mol$^{-1}$ ; $H_1^{E,\infty}$ (aniline + 2-propanone)[22] = −5.68 and $H_1^{E,\infty}$ (aniline + heptane)[27] = 15 kJ•mol$^{-1}$. Therefore, $\Delta H_{\text{(N-CO)bond}}$

$= -30.52$ kJ•mol$^{-1}$. This value is even lower than that for the H-bonds between 1-alkanol molecules, which in the ERAS model is assumed to be[27,28] $-25.1$ kJ•mol$^{-1}$.

Mixtures such as amine + 1-alcohol, or + CHCl$_3$, characterized also by strong interactions between unlike molecules,[28,29] show similar $V^E$ values to those given in Table 2. So, $V^E$ (1-propanol + propylamine)[30] $= -1.315$ cm$^3$•mol$^{-1}$; and $V^E$ (trichloromethane + butylamine)[31] $= -0.368$ cm$^3$•mol$^{-1}$. The negative $\left(\frac{\partial V^E}{\partial T}\right)_P$ and $\kappa_S^E$ values are in agreement with the existence of strong ketone-amine interactions in the investigated systems. The former have been interpreted in terms of a decrease in the molar volume of complex formation, which overcompensates for the decrease in the extent of complex formation, and have been encountered, e.g., in amine + trichloromethane mixtures.[31,32]

On the other hand, it is remarkable that the composition dependence of the $V^E$ and $\kappa_S^E$ curves is similar and is the same for the studied mixtures (Figures 1 and 2). These curves are shifted towards higher mole fractions of 2-propanone (the smaller component), and show a minimum at $x_1 = 0.6$. The fact that the $H^E$ curve of the 2-propanone + aniline system also shows a minimum at $x_1 = 0.55$ suggests that structural effects are rather weak.

Finally, it is noticeable that $V^E$ and $\kappa_S^E$ increase in the same sequence: aniline < N-methylaniline < pyridine (Figures 1 and 2). This may be interpreted assuming that the new 2-propanone-amine interactions created upon mixing are more easily formed in the case of aniline solutions due to the larger ability of aniline to form H-bonds with the oxygen atom of the ketone, related with the presence of the NH$_2$ group in this amine.

**Conclusions**

In this work, we have determined $V^E$, $u^E$, $\alpha_P^E$ and $\kappa_S^E$. The $V^E$ and $\kappa_S^E$ magnitudes are negative and increase in the sequence: aniline < N-methylaniline < pyridine. The corresponding curves are shifted towards higher mole fractions of 2-propanone. The data have been interpreted in terms of acetone-amine interactions and weak structural effects. The existence of strong 2-propanone-amine interactions is supported by negative $(\partial V^E / \partial T)_P$ values, and by the enthalpy of 2-propanone-aniline association which is more negative than that of the H-bonds between 1-alcohol molecules.

**ACKNOWLEDGEMENTS**

The authors gratefully acknowledge the financial support received from the Consejería de Educación y Cultura of Junta de Castilla y León, under Projects VA075A07 and VA052A09 and from the Ministerio de Educación y Ciencia, under the Project FIS2007-61833. I.A. and V.A. also gratefully acknowledge the grants received from the Junta de Castilla y León.

**TABLE 1**

Physical Properties of Pure Compounds, 2-propanone, Aniline, *N*-methylaniline and Pyridine at temperature *T*.

| Property | $T$/K | 2-propanone | | Aniline | | *N*-methylaniline | | Pyridine | |
|---|---|---|---|---|---|---|---|---|---|
| | | This work | Lit | This work | Lit | This work | Lit. | This work | Lit. |
| $\rho$/g•cm$^3$ | 293.15 | 0.790694 | 0.78998[a] | 1.021702 | 1.02104[b] | 0.986077 | | 0.983053 | 0.98319[a] |
| | 298.15 | 0.785320 | 0.784431[c] | 1.017406 | 1.01710[b] | 0.982066 | 0.98206[d] | 0.978050 | 0.97824[a] |
| | | | 0.78428[e] | | 1.01744[d] | | | | 0.97810[f] |
| | | | 0.78457[g] | | 1.01741[h] | | | | |
| | 303.15 | 0.778812 | 0.77914[e] | 1.013045 | 1.01284[b] | 0.978011 | | 0.972980 | 0.97286[i] |
| $u$ /m•s$^{-1}$ | 293.15 | 1182.8 | 1192[j] | 1657.0 | 1651.3[k] | 1582.3 | | 1436.7 | |
| | 298.15 | 1160.7 | 1161.72[c] | 1638.6 | 1634[h] | 1564.4 | | 1416.7 | |
| | | | 1154[e] | | 1632.8[k] | | | | |
| | | | 1160.6[g] | | | | | | |
| | 303.15 | 1138.7 | 1132.2[e] | 1619.2 | 1614.5[k] | 1545.5 | | 1396.6 | 1398[i] |
| $\alpha_P$/10$^{-3}$K$^{-1}$ | 298.15 | 1.56 | 1.426[e] | 0.85 | 0.850[a] | 0.82 | | 1.03 | 1.07[a] |
| $\kappa_S$/TPa$^{-1}$ | 293.15 | 903.9 | | 356.5 | | 405.0 | | 492.8 | |
| | 298.15 | 945.1 | 944.59[c] | 366.1 | 368[a] | 416.1 | | 509.5 | |
| | | | 958[e] | | | | | | |
| | | | 946[g] | | | | | | |
| | 303.15 | 990.4 | 1003[e] | 376.5 | | 428.1 | | 526.9 | 525[i] |
| $\kappa_T$/TPa$^{-1}$ | 298.15 | 1373.6 | 1324[a] | 467.9 | 472[b] | 522.1 | | 704 | |
| | | | 1330[e] | | | | | | |
| $C_P$/ J•mol$^{-1}$•K$^{-1}$ | 298.15 | | 124.9[a] | | 194.1[l] | | 207.1[l] | | 131.5[m] |

$\rho$, density; $u$, speed of sound; $\alpha_P$, isobaric thermal expansion coefficient; $\kappa_S$, adiabatic compressibility; $\kappa_T$, isothermal compressibility (calculated from $\kappa_T = \kappa_S + \dfrac{TV\alpha_P^2}{C_P}$) and $C_P$, isobaric heat capacity.

[a]Ref. 5; [b]Ref. 33; [c]Ref. 21; [d]Ref. 34; [e]Ref. 35; [f]Ref. 36; [g]Ref. 37; [h] Ref. 38; [i]Ref. 39; [j]Ref. 40; [k] Ref. 41; [l]Ref. 42; [m]Ref. 20

**TABLE 2**

Densities, $\rho$, Molar Excess Volumes, $V^E$, and Deviations from the Ideal Behaviour of the Speed of Sound for 2-propanone(1) + Aromatic Amine(2) Mixtures at Temperature T.

| $x_1$ | $\rho$/g•cm$^{-3}$ | $V^E$ / cm$^3$•mol$^{-1}$ | $u$ /m•s$^{-1}$ | $x_1$ | $\rho$/g•cm$^{-3}$ | $V^E$ / cm$^3$•mol$^{-1}$ | $u$ /m•s$^{-1}$ |
|---|---|---|---|---|---|---|---|
| \multicolumn{8}{c}{2-propanone(1) + aniline(2) ; T/K = 293.15} |
| 0.0522 | 1.013709 | −0.1611 | 1638.35 | 0.5497 | 0.920272 | −1.160 | 1437.53 |
| 0.1078 | 1.004965 | −0.3316 | 1618.48 | 0.5830 | 0.912529 | −1.169 | 1421.64 |
| 0.1567 | 0.996941 | −0.4681 | 1600.65 | 0.6446 | 0.897535 | −1.153 | 1390.65 |
| 0.2074 | 0.988429 | −0.6142 | 1581.82 | 0.6911 | 0.885684 | −1.118 | 1366.49 |
| 0.2490 | 0.981269 | −0.7314 | 1566.20 | 0.7465 | 0.870629 | −1.024 | 1335.85 |
| 0.3016 | 0.971707 | −0.8513 | 1545.76 | 0.8046 | 0.854022 | −0.8819 | 1302.73 |
| 0.3454 | 0.963326 | −0.9303 | 1527.88 | 0.8497 | 0.840545 | −0.7433 | 1276.21 |
| 0.3832 | 0.955799 | −0.9842 | 1512.13 | 0.9045 | 0.823238 | −0.5220 | 1242.96 |
| 0.4313 | 0.946013 | −1.052 | 1491.44 | 0.9475 | 0.809118 | −0.3046 | 1216.10 |
| 0.4908 | 0.933416 | −1.120 | 1465.03 | | | | |
| \multicolumn{8}{c}{2-propanone(1) + aniline(2); T/K = 298.15} |
| 0.0569 | 1.008670 | −0.1850 | 1618.3 | 0.5501 | 0.915531 | −1.203 | 1417.8 |
| 0.1088 | 1.000436 | −0.3486 | 1599.6 | 0.5922 | 0.905703 | −1.215 | 1397.3 |
| 0.1539 | 0.993118 | −0.4898 | 1583.2 | 0.6499 | 0.891425 | −1.190 | 1367.7 |
| 0.1978 | 0.985694 | −0.6111 | 1566.7 | 0.6993 | 0.878643 | −1.145 | 1341.6 |
| 0.2484 | 0.976895 | −0.7492 | 1547.4 | 0.7522 | 0.864125 | −1.048 | 1312.1 |
| 0.3039 | 0.966770 | −0.8805 | 1525.8 | 0.8082 | 0.848052 | −0.9146 | 1279.9 |
| 0.3505 | 0.957976 | −0.9826 | 1507.0 | 0.8533 | 0.834272 | −0.7530 | 1252.8 |
| 0.4114 | 0.945773 | −1.081 | 1481.1 | 0.8951 | 0.821048 | −0.5816 | 1227.3 |
| 0.4585 | 0.935933 | −1.142 | 1460.4 | 0.9545 | 0.801258 | −0.2750 | 1190.0 |
| 0.5035 | 0.926122 | −1.181 | 1439.9 | | | | |
| \multicolumn{8}{c}{2-propanone(1) + aniline(2); T/K = 303.15} |
| 0.0515 | 1.005017 | −0.1691 | 1600.76 | 0.5502 | 0.910191 | −1.261 | 1397.55 |
| 0.1002 | 0.997272 | −0.3311 | 1583.36 | 0.6037 | 0.897422 | −1.269 | 1371.13 |
| 0.1475 | 0.989506 | −0.4805 | 1566.03 | 0.6464 | 0.886697 | −1.246 | 1348.94 |
| 0.1926 | 0.981799 | −0.6104 | 1549.06 | 0.6975 | 0.873362 | −1.197 | 1321.68 |
| 0.2399 | 0.973527 | −0.7470 | 1531.02 | 0.7548 | 0.857721 | −1.111 | 1290.18 |
| 0.3024 | 0.962003 | −0.8977 | 1506.37 | 0.8078 | 0.842147 | −0.9626 | 1259.19 |
| 0.3475 | 0.953088 | −0.9716 | 1487.41 | 0.8567 | 0.827148 | −0.7897 | 1229.86 |
| 0.3947 | 0.943989 | −1.087 | 1468.09 | 0.9014 | 0.812779 | −0.5938 | 1202.26 |
| 0.4503 | 0.932550 | −1.183 | 1444.06 | 0.9494 | 0.796575 | −0.3342 | 1171.73 |
| 0.4979 | 0.922075 | −1.225 | 1422.10 | | | | |
| 2-propanone(1) + N-Methylaniline(2); $T$/K = 293.15 | | | | | | | |
| 0.0565 | 0.979562 | −0.1177 | 1566.00 | 0.5489 | 0.905091 | −0.6971 | 1396.72 |
| 0.1032 | 0.973934 | −0.2094 | 1552.20 | 0.5965 | 0.895722 | −0.6994 | 1377.18 |
| 0.1494 | 0.968145 | −0.2955 | 1538.20 | 0.6485 | 0.884852 | −0.6849 | 1354.94 |
| 0.1916 | 0.962605 | −0.3664 | 1524.93 | 0.7018 | 0.873046 | −0.6508 | 1331.41 |
| 0.2510 | 0.954413 | −0.4576 | 1505.74 | 0.7472 | 0.862270 | −0.5948 | 1310.35 |
| 0.2952 | 0.948053 | −0.5209 | 1491.11 | 0.8020 | 0.848690 | −0.5245 | 1284.63 |
| 0.3555 | 0.938912 | −0.5953 | 1470.34 | 0.8563 | 0.834277 | −0.4247 | 1258.05 |
| 0.4039 | 0.931040 | −0.6344 | 1452.78 | 0.9053 | 0.820343 | −0.3047 | 1233.13 |
| 0.4478 | 0.923576 | −0.6619 | 1436.33 | 0.9548 | 0.805391 | −0.1564 | 1207.00 |
| 0.4981 | 0.914659 | −0.6888 | 1417.03 | | | | |
| 2-propanone(1) + N-Methylaniline(2); $T$/K = 298.15 | | | | | | | |
| 0.0440 | 0.977005 | −0.0971 | 1551.80 | 0.5563 | 0.899086 | −0.7343 | 1374.44 |
| 0.0984 | 0.970455 | −0.2105 | 1535.67 | 0.6004 | 0.890231 | −0.7371 | 1355.78 |
| 0.1447 | 0.964607 | −0.2985 | 1521.51 | 0.6473 | 0.880387 | −0.7266 | 1335.71 |
| 0.1996 | 0.957353 | −0.3971 | 1504.22 | 0.6962 | 0.869434 | −0.6904 | 1313.74 |
| 0.2503 | 0.950290 | −0.4773 | 1487.65 | 0.7506 | 0.856501 | −0.6268 | 1288.41 |
| 0.2919 | 0.944327 | −0.5456 | 1474.09 | 0.8009 | 0.843758 | −0.5449 | 1264.11 |
| 0.3511 | 0.935283 | −0.6199 | 1453.29 | 0.8518 | 0.830205 | −0.4516 | 1238.96 |
| 0.3924 | 0.928532 | −0.6516 | 1438.13 | 0.8986 | 0.816915 | −0.3377 | 1215.04 |
| 0.4520 | 0.918415 | −0.6986 | 1415.84 | 0.9488 | 0.801799 | −0.1884 | 1188.58 |
| 0.4948 | 0.910794 | −0.7256 | 1399.32 | | | | |
| 2-propanone(1) + N-Methylaniline(2); $T$/K = 303.15 | | | | | | | |
| 0.0461 | 0.972638 | −0.1079 | 1532.14 | 0.5507 | 0.895228 | −0.7946 | 1357.06 |
| 0.0934 | 0.966889 | −0.2135 | 1518.12 | 0.5949 | 0.886274 | −0.7909 | 1338.42 |
| 0.1524 | 0.959343 | −0.3337 | 1500.03 | 0.6467 | 0.875210 | −0.7715 | 1315.78 |
| 0.1937 | 0.953815 | −0.4108 | 1486.97 | 0.6928 | 0.864807 | −0.7384 | 1295.01 |
| 0.2442 | 0.946596 | −0.4829 | 1470.07 | 0.7424 | 0.852961 | −0.6836 | 1271.81 |
| 0.2834 | 0.941069 | −0.5653 | 1457.52 | 0.7993 | 0.838524 | −0.5956 | 1244.29 |
| 0.3357 | 0.933026 | −0.6316 | 1439.18 | 0.8469 | 0.825646 | −0.4966 | 1220.47 |
| 0.3991 | 0.922737 | −0.7029 | 1416.25 | 0.8977 | 0.811117 | −0.3691 | 1194.30 |

TABLE 2 (continued)

| | | | | | | | |
|---|---|---|---|---|---|---|---|
| 0.4460 | 0.914773 | −0.7526 | 1399.25 | 0.9484 | 0.795598 | −0.2065 | 1167.27 |
| 0.4904 | 0.906677 | −0.7704 | 1381.24 | | | | |

2-propanone(1) + pyridine(2);  $T/K = 293.15$

| | | | | | | | |
|---|---|---|---|---|---|---|---|
| 0.0577 | 0.973393 | −0.0418 | 1424.88 | 0.5519 | 0.884082 | −0.2351 | 1309.32 |
| 0.1007 | 0.966028 | −0.0648 | 1415.57 | 0.5969 | 0.875296 | −0.2332 | 1297.67 |
| 0.1536 | 0.956946 | −0.0993 | 1404.09 | 0.6425 | 0.866297 | −0.2290 | 1285.63 |
| 0.2043 | 0.948163 | −0.1319 | 1392.96 | 0.7030 | 0.854168 | −0.2181 | 1269.32 |
| 0.2219 | 0.945011 | −0.1369 | 1389.00 | 0.7504 | 0.844514 | −0.2028 | 1256.33 |
| 0.2968 | 0.931713 | −0.1766 | 1371.80 | 0.8066 | 0.832904 | −0.1802 | 1240.56 |
| 0.3537 | 0.921354 | −0.1952 | 1358.39 | 0.8569 | 0.822218 | −0.1428 | 1225.97 |
| 0.4051 | 0.911899 | −0.2117 | 1346.07 | 0.9019 | 0.812518 | −0.1041 | 1212.66 |
| 0.5191 | 0.890383 | −0.2319 | 1317.70 | 0.9456 | 0.803006 | −0.0652 | 1199.55 |

2-propanone(1) + pyridine(2);  $T/K = 298.15$

| | | | | | | | |
|---|---|---|---|---|---|---|---|
| 0.0514 | 0.969365 | −0.0338 | 1405.83 | 0.5448 | 0.880262 | −0.2568 | 1290.28 |
| 0.0969 | 0.961623 | −0.0661 | 1396.12 | 0.5940 | 0.870649 | −0.2570 | 1277.31 |
| 0.1486 | 0.952705 | −0.0997 | 1384.77 | 0.6454 | 0.860492 | −0.2535 | 1263.72 |
| 0.2023 | 0.943375 | −0.1356 | 1372.83 | 0.6971 | 0.850006 | −0.2349 | 1249.50 |
| 0.2444 | 0.935930 | −0.1598 | 1363.28 | 0.7518 | 0.838783 | −0.2125 | 1234.34 |
| 0.2960 | 0.926692 | −0.1850 | 1351.30 | 0.8049 | 0.827707 | −0.1844 | 1219.28 |
| 0.3544 | 0.916021 | −0.2062 | 1337.42 | 0.8456 | 0.819096 | −0.1575 | 1207.46 |
| 0.3965 | 0.908330 | −0.2268 | 1327.36 | 0.8955 | 0.808385 | −0.1177 | 1192.71 |
| 0.4516 | 0.898006 | −0.2393 | 1313.79 | 0.9456 | 0.797415 | −0.0666 | 1177.52 |
| 0.5036 | 0.888125 | −0.2488 | 1300.66 | | | | |

2-propanone(1) + pyridine(2);  $T/K = 303.15$

| | | | | | | | |
|---|---|---|---|---|---|---|---|
| 0.0493 | 0.964637 | −0.0401 | 1386.22 | 0.4970 | 0.883593 | −0.2683 | 1281.42 |
| 0.0954 | 0.956714 | −0.0743 | 1376.21 | 0.5412 | 0.875069 | −0.2759 | 1270.11 |
| 0.1463 | 0.947874 | −0.1106 | 1365.00 | 0.5965 | 0.864243 | −0.2804 | 1255.54 |
| 0.1983 | 0.938736 | −0.1457 | 1353.35 | 0.6594 | 0.851677 | −0.2746 | 1238.78 |
| 0.2424 | 0.930863 | −0.1707 | 1343.26 | 0.6912 | 0.845142 | −0.2607 | 1229.99 |
| 0.2929 | 0.921779 | −0.2001 | 1331.54 | 0.7491 | 0.833115 | −0.2306 | 1213.69 |
| 0.3522 | 0.910875 | −0.2233 | 1317.33 | 0.8475 | 0.812305 | −0.1697 | 1185.30 |
| 0.4048 | 0.901126 | −0.2463 | 1304.51 | 0.8946 | 0.802071 | −0.1259 | 1171.30 |
| 0.4497 | 0.892629 | −0.2575 | 1293.35 | 0.9443 | 0.791125 | −0.0727 | 1156.13 |

**TABLE 3**

Excess functions at 298.15 K for $\kappa_S$, Adiabatic Compressibility, $u$, Speed of Sound, and $\alpha_P$, Isobaric Thermal Expansion Coefficient of 2-propanone(1) + Aromatic Amine(2) Mixtures.

| $x_1$ | $\kappa_S^E$ /TPa$^{-1}$ | $u^E$ /m·s$^{-1}$ | $\alpha_P^E$ /10$^{-6}$·K$^{-1}$ |
|---|---|---|---|
| | 2-propanone(1) + aniline(2) | | |
| 0.0569 | −20.43 | 40.4 | −8.84 |
| 0.1088 | −38.63 | 70.7 | −17.60 |
| 0.1539 | −53.97 | 92.6 | −25.78 |
| 0.1978 | −68.26 | 110.2 | −34.13 |
| 0.2484 | −83.96 | 126.6 | −44.26 |
| 0.3039 | −100.14 | 140.4 | −55.68 |
| 0.3505 | −112.54 | 148.5 | −65.46 |
| 0.4114 | −126.55 | 154.4 | −77.89 |
| 0.4585 | −135.71 | 156.0 | −87.00 |
| 0.5035 | −142.74 | 155.0 | −94.91 |
| 0.5501 | −147.91 | 151.5 | −101.96 |
| 0.5922 | −150.71 | 146.3 | −107.09 |
| 0.6499 | −150.56 | 135.7 | −111.20 |
| 0.6993 | −146.83 | 124.2 | −111.56 |
| 0.7522 | −137.58 | 108.6 | −107.53 |
| 0.8082 | −121.87 | 89.2 | −97.67 |
| 0.8533 | −103.18 | 71.1 | −84.25 |
| 0.8951 | −81.16 | 52.8 | −67.19 |
| 0.9545 | −40.17 | 24.1 | −33.77 |
| | 2-propanone(1) + N-methylaniline(2) | | |
| 0.0440 | −10.58 | 18.3 | −6.69 |
| 0.0984 | −23.36 | 38.0 | −15.97 |
| 0.1447 | −33.93 | 52.4 | −24.49 |
| 0.1996 | −46.09 | 67.1 | −35.09 |
| 0.2503 | −56.75 | 78.2 | −45.01 |
| 0.2919 | −65.36 | 86.1 | −53.12 |
| 0.3511 | −76.19 | 94.0 | −64.07 |
| 0.3924 | −82.84 | 97.7 | −71.02 |
| 0.4520 | −91.51 | 101.1 | −79.97 |
| 0.4948 | −96.68 | 101.8 | −85.28 |

TABLE 3 (continued)

| | | | |
|---|---|---|---|
| 0.5563 | − 101.86 | 100.2 | − 90.63 |
| 0.6004 | − 103.80 | 97.0 | − 92.68 |
| 0.6473 | − 104.40 | 92.4 | − 92.94 |
| 0.6962 | − 102.29 | 85.5 | − 90.62 |
| 0.7506 | − 96.46 | 75.6 | − 84.65 |
| 0.8009 | − 87.21 | 64.3 | − 75.58 |
| 0.8518 | − 73.81 | 51.1 | − 62.79 |
| 0.8986 | − 56.69 | 37.0 | − 47.23 |
| 0.9488 | − 32.34 | 19.8 | − 26.30 |
| | 2-propanone(1) + pyridine(2) | | |
| 0.0514 | − 9.99 | 13.0 | − 2.43 |
| 0.0969 | − 18.39 | 22.9 | − 6.47 |
| 0.1486 | − 27.19 | 32.3 | − 12.46 |
| 0.2023 | − 35.60 | 40.4 | − 19.63 |
| 0.2444 | − 41.61 | 45.5 | − 25.47 |
| 0.2960 | − 48.11 | 50.3 | − 32.38 |
| 0.3544 | − 54.32 | 54.1 | − 39.36 |
| 0.3965 | − 58.19 | 56.0 | − 43.68 |
| 0.4516 | − 61.95 | 56.9 | − 47.80 |
| 0.5036 | − 64.26 | 56.6 | − 50.09 |
| 0.5448 | − 65.42 | 55.7 | − 50.73 |
| 0.5940 | − 65.25 | 53.3 | − 49.91 |
| 0.6454 | − 63.91 | 50.1 | − 47.38 |
| 0.6971 | − 60.56 | 45.5 | − 43.00 |
| 0.7518 | − 55.39 | 39.8 | − 36.84 |
| 0.8049 | − 48.27 | 33.2 | − 29.58 |
| 0.8456 | − 41.20 | 27.4 | − 23.44 |
| 0.8955 | − 30.58 | 19.5 | − 15.57 |
| 0.9456 | − 17.40 | 10.6 | − 7.72 |

**TABLE 4**

Coefficients $A_i$ and Standard Deviations, $\sigma(F^E)$ (eq. 8) for Representation of the $F^{E,a}$ Property at Temperature $T$ for 2-propanone(1) + Aromatic Amine(2) Systems by eq. 7

| System[b] | T/K | Property $F^E$ | $A_0$ | $A_1$ | $A_2$ | $A_3$ | $\sigma(F^E)$ |
|---|---|---|---|---|---|---|---|
| 2-propanone + aniline | 293.15 | $V^E$ | −4.541 | −1.58 | −0.34 | | 0.008 |
| | 298.15 | $V^E$ | −4.731 | −1.614 | −0.31 | | 0.005 |
| | | $u^E$ | 620.66 | −88.3 | 40.7 | −28.6 | 0.10 |
| | | $\kappa_S^E$ | −569.04 | −278.8 | −98.9 | −27.8 | 0.09 |
| | | $\alpha_P^E$ | −377.64 | −337.6 | −113.8 | | 0.10 |
| | 303.15 | $V^E$ | −4.905 | −1.85 | −0.42 | | 0.008 |
| 2-propanone + N-Methylaniline | 293.15 | $V^E$ | −2.760 | −0.754 | −0.22 | | 0.004 |
| | 298.15 | $V^E$ | −2.908 | −0.801 | −0.19 | | 0.004 |
| | | $u^E$ | 407.00 | −13.5 | 14.8 | | 0.11 |
| | | $\kappa_S^E$ | −388.4 | −204.4 | −83.9 | −31 | 0.14 |
| | | $\alpha_P^E$ | −343.32 | −211.97 | −10.1 | | 0.05 |
| | 303.15 | $V^E$ | −3.106 | −0.907 | −0.20 | | 0.005 |
| 2-propanone + pyridine | 293.15 | $V^E$ | −0.922 | −0.271 | −0.096 | | 0.002 |
| | 298.15 | $V^E$ | −1.003 | −0.296 | | | 0.002 |
| | | $u^E$ | 226.71 | −32.2 | 11.9 | | 0.07 |
| | | $\kappa_S^E$ | −256.76 | −69.8 | −18.5 | −5.3 | 0.08 |
| | | $\alpha_P^E$ | −200.09 | −59.6 | 125.0 | 6.1 | 0.04 |
| | 303.15 | $V^E$ | −1.090 | −0.317 | | | 0.003 |

[a] $F^E = V^E$, units: cm³·mol⁻¹; $F^E = u^E$, units: m·s⁻¹; $F^E = \kappa_S^E$, units: TPa⁻¹; $F^E = \alpha_P^E$, units: 10⁻⁶·K⁻¹

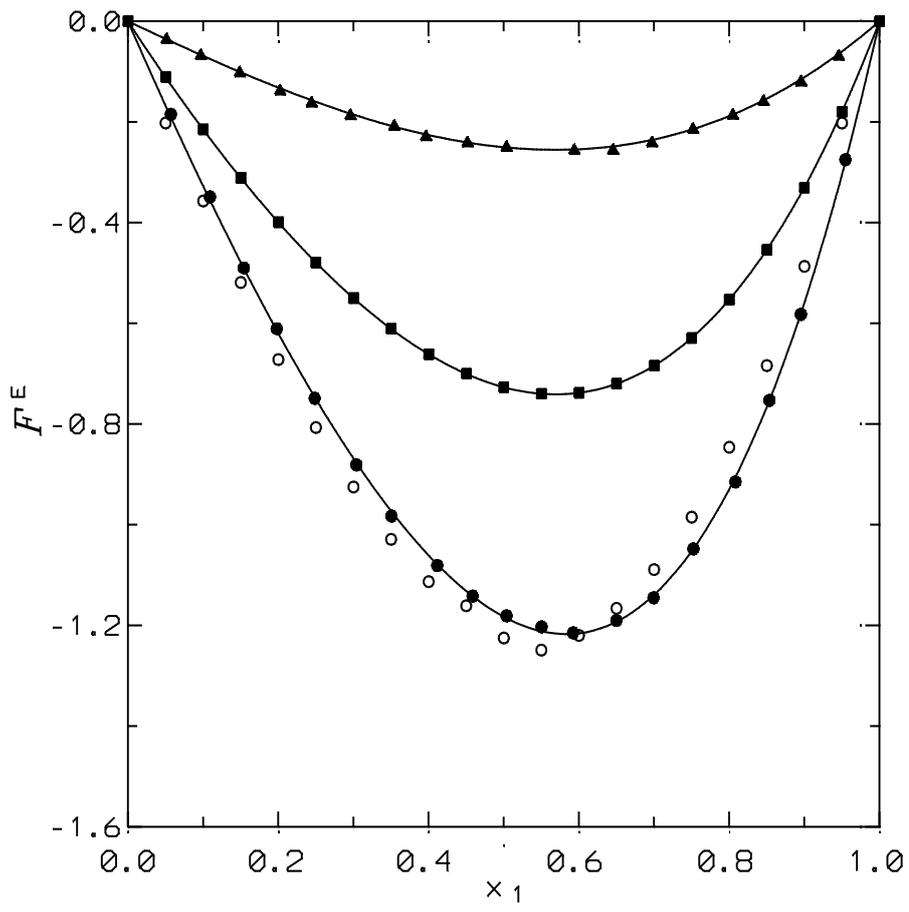

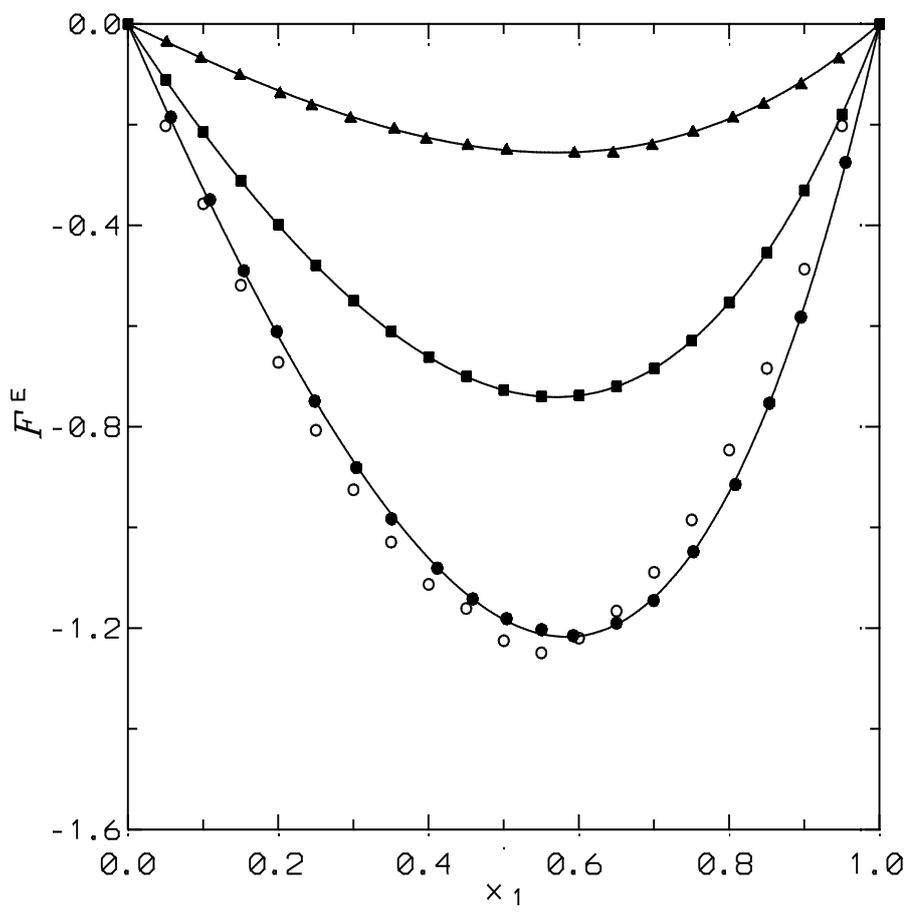

Figure 1  Excess molar functions for the 2-propanone(1) + aromatic amine(2) systems at atmospheric pressure and 298.15 K. Full symbols,  $V^E$ /cm$^3$•mol$^{-1}$ (this work): ●, aniline; ■, $N$-methylaniline; ▲, pyridine. Open symbols, ($H^E$/J•mol$^{-1}$)/100 for the aniline solution.[21] Solid lines, calculations with eq. 7 using the coefficients from Table 4.

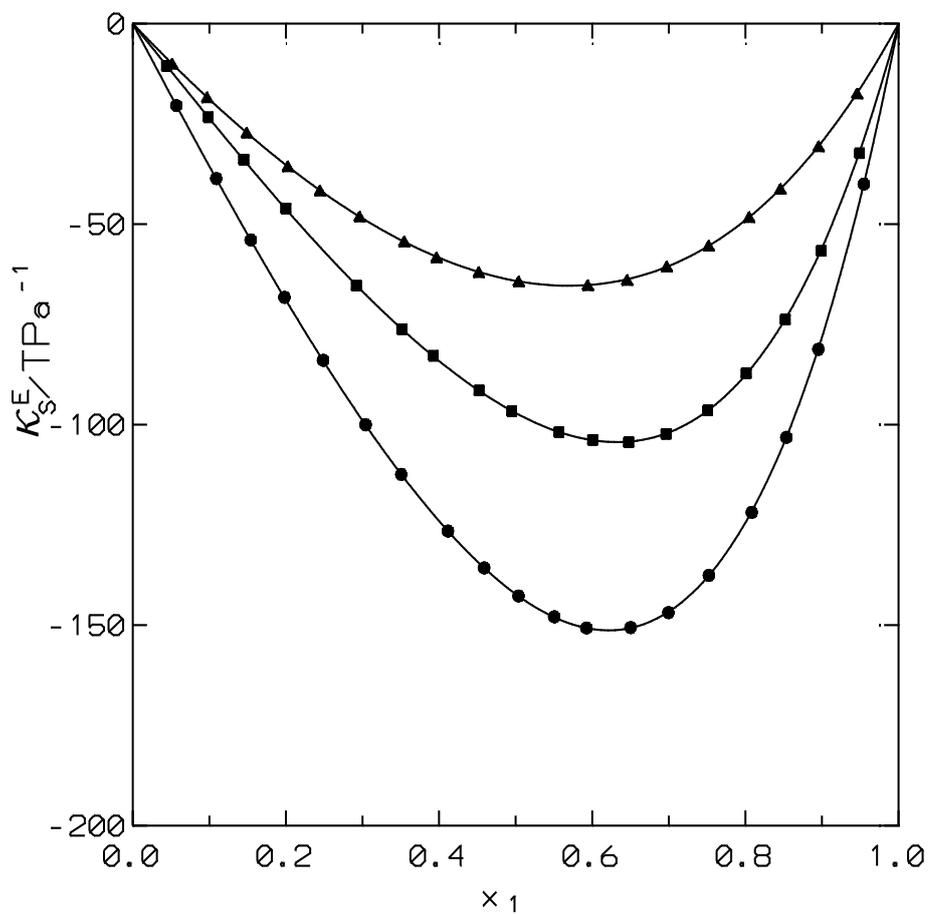

Figure 2. $\kappa_S^E$ for the 2-propanone(1) + aromatic amine(2) systems at atmospheric pressure and 298.15 K. Symbols, experimental data (this work): ●, aniline; ■, *N*-methylaniline; ▲, pyridine. Solid lines, calculations with eq. 7 using the coefficients from Table 4.